\documentclass[12pt]{article}

\usepackage{graphicx}
\usepackage{amsmath}
\usepackage{lmodern}
\usepackage{comment}
\usepackage{mathtools}
\usepackage{adjustbox}
\usepackage{hyperref}
\usepackage{amsfonts}
\usepackage[letterpaper,margin=1in]{geometry}
\usepackage{url}

\renewcommand{\vec}[1]{\mathbf{#1}}

\renewenvironment{abstract}
	{\quotation}
	{\endquotation}
\date{}

\makeatletter
\renewcommand{\fnum@figure}{\textbf{Figure \thefigure}}
\renewcommand{\fnum@table}{\textbf{Table \thetable}}
\makeatother

\title{\bfseries \boldmath Physics-informed denoising method for image reconstruction in quantitative low-field MRI}

\author{
	Catarina Redshaw Kranich$^{1\ast}$,
	Claudia Prieto$^{2}$,
    Christoph Kolbitsch$^{1}$\and
    Felix Frederik Zimmermann$^{1}$\and
	\small$^{1}$Physikalisch-Technische Bundesanstalt, Braunschweig and Berlin, Germany\and\
	\small$^{2}$Pontificia Universidad Católica de Chile, Santiago de Chile, Chile\and
	\small$^\ast$Corresponding author. Email: catarina.redshaw-kranich@ptb.de\and
}

\begin{document} 

\maketitle

\begin{abstract}\bfseries \boldmath
Low-field magnetic resonance imaging (MRI) is becoming increasingly important for medical imaging because it can reduce healthcare costs while ensuring high diagnostic output. Nevertheless, quantitative imaging in low-field MRI faces challenges, such as low signal-to-noise ratio and long scan durations. Deep learning approaches have been proposed for image reconstruction to overcome these challenges. Still, deep learning often requires large high-quality training datasets which are usually not available for low-field applications. Here we propose a modular unrolled end-to-end deep learning method for the denoised reconstruction of quantitative parameter maps directly from k-space data for low-field MRI.  It consists of three sub-networks that are iteratively applied. They are used for the regularization of the quantitative parameter estimation, as well as for the signal estimation that is based on simulated signal curves. It generalises well and can be applied to different field strengths and even different quantitative MR sequences without the need for new training data. We applied the presented method to noisy data of knees acquired at $0.55\,$T for the reconstruction of $T_2$-maps and compared it to other classical and deep learning methods. We also applied the proposed approach to $T_1$-mapping of knees at $72\,$mT and  $T_2$-mapping of brains at $0.6\,$T. The presented approach outperforms the other reconstruction methods with a median difference below $4\,$ms to the ground truth $T_2$-map. Even though the network was trained with $T_2$-maps acquired at $0.55\,$T, it successfully denoised data acquired at different field strengths, sequences, and of different anatomies. As a result, the proposed network and its underlying method offer an efficient and flexible solution to denoise low-field MR data and make quantitative low-field MRI a feasible diagnostic tool for clinical applications. 
\end{abstract}
\section{Introduction}\label{sec:introduction}

In recent years low-field magnetic resonance imaging (MRI) ($<1$\,T) \cite{Kofler2026} has become more popular, offering benefits like reduced costs and fewer safety concerns \cite{Arnold2023}.  The clinical applications for low-field MRI include many different applications, such as qualitative neuro imaging \cite{Liu2021} or orthopedic  imaging \cite{Pogarell2024}. In contrast to qualitative MRI, quantitative MRI (qMRI) offers the opportunity to measure (bio-) physical tissue properties, such as the relaxation times $T_1$ or $T_2$ \cite{Gulani2020}. For this mapping of quantitative parameters, commonly data acquisition has to be repeated with varying acquisition parameters (e.g. different echo times for $T_2$-mapping). Quantitative scans of patients often require longer scan times than qualitative imaging.  One of the main challenges of low-field MRI is the low signal-to-noise ratio (SNR) \cite{Webb2023} which often can also only be addressed by increasing scan times, making quantitative imaging at low-field even more challenging to achieve in a clinically feasible scan time. \\
Classical approaches have been used to address this problem, such as for low-field MRI with image-based and k-space based low rank reconstruction \cite{Dong2025}. To improve the accuracy of quantitative parameter estimation, Principal Component Analysis (PCA) Compression has been used for higher magnetic field strengths (3\,T) \cite{Pfister2019}, where qualitative images are mapped onto a temporal subspace with lower dimension.\\
Deep learning approaches have been shown to outperform classical approaches also for low-field MRI \cite{Koonjoo2021,Salehi2025,deLeeuwdenBouter2022}. They can be used as a post-processing step applied to reconstructed images \cite{Xue2025,Hauptmann2019, Zimmermann2025_2}. Even though this improves the image quality, it does not ensure data consistency. End-to-end networks can enforce this data consistency \cite{ Aggarwal2019} and have been shown to outperform post-processing methods \cite{ Kofler2024}. The main challenge of applying these networks to  low-field MRI is that they  require large amounts of high-quality training data which is difficult to obtain for low-field MRI.  Training data acquired at high fields could be used but would lead to a generalization gap when applied to low-field data. \\
To overcome this challenge we propose a deep learning based approach which generalises very well across different field strengths, anatomies and quantitative parameters. It does not require to be retrained with domain specific high-quality training data. The proposed approach is a novel unrolled end-to-end deep learning solution to directly reconstruct quantitative maps from k-space data. It consists of different networks, where one of them is used as a sequence-agnostic regularization on the quantitative maps. Another anatomy-agnostic network is pre-trained on simulated signal curves to incorporate complex quantitative signal models, such as extended phase graphs. This network works on a pixel-by-pixel basis and can be easily adapted to different parameters without having to re-train the entire method. The proposed approach was trained and evaluated on $T_2$-mapping data of the knee obtained on a 0.55\,T Siemens MAGNETOM Free.Max. We demonstrated transfer to unseen anatomies (brain, 0.6\,T) and unseen field strengths and sequences ($T_1$-mapping, 72\,mT).

\section{Methodology}\label{sec:methodology}
The presented approach reconstructs quantitative maps directly from acquired k-space data. First, the general problem in qMRI reconstruction is presented. Afterwards the different components of the proposed network are explained. 
\subsection{qMRI Reconstruction Problem }\label{sec:problem}
In qMRI k-space data $\vec{k}\in\mathbb{C}^{N_a\cdot{N_c\cdot N_k}}$ is acquired to determine parameter maps for $N_p$ different quantitative parameters $\vec{p}=[\vec{p}_{1},\vec{p}_{2},...,\vec{p}_{N_p}]^{T}$ with $\vec{p}\in\mathcal{P} \coloneqq \mathcal{P}_{1} \times\cdots\times\mathcal{P}_{N_p}$  and $\mathcal{P}_p\in\{{\mathbb{R}^{N},\mathbb{C}^{N}}\},p=1,\cdots,N$. $N_a$ is the number of different acquisitions that have to be made in order to deduce quantitative information from the qualitative images $\vec{x}\in\mathbb{C}^{N_a\cdot N}$ with $N$ voxels. $N_c$ is the number of receiver coils and $N_k$ denotes the number of acquired k-space points. The forward problem for qMRI is given by
\begin{equation}
    \vec{k} = \left(\vec{E}\circ q\right)\left(\vec{p}\right) + \epsilon
\end{equation}
where $\vec{E}:\mathbb{C}^{N\cdot N_a}\rightarrow\mathbb{C}^{N_{c}\cdot N_{k}\cdot N_a}$ is the encoding operator. $q$ denotes the signal model, that is commonly non-linear, and $\epsilon$ stands for random Gaussian noise. The signal model $q:\mathcal{P}\rightarrow\mathbb{C}^{N_a\cdot N}$ maps the parameters $\vec{p}$ to qualitative images $\vec{x}\in \mathbb{C}^{N_a\cdot N}$ according to different sequence settings . $\vec{E}$ can be written as $\vec{E}=\vec{SFC}$, where $\vec{S}$ denotes a diagonal sampling mask, $\vec{F}$ is  the Fourier operator and $\vec{C}$ the coil sensitivity operator.  $\vec{S}$ can either be fully-sampled ($S_{m}=1$) or undersampled ($S_{m} \in \{0,1\}$) with  $m=1,...,N_{k}$. The reconstruction problem to obtain the parameter vector $\vec{p}$ from the acquired k-space data $\vec{k}$ is commonly ill-posed due to the noise $\epsilon$ and potential undersampling. To obtain a robust solution, regularization needs to be employed:
\begin{equation}
    \text{arg}\min_{\vec{p}}||\vec{SFC}q\left(\vec{p}\right)-\vec{k}||^{2}_{2}+\mathcal{R}\left(\vec{p}\right)
    \label{eq:inverse_problem}
\end{equation}
$\mathcal{R}$ denotes the regularization term constraining the parameter vectors $\vec{p}$.\cite{Kofler2024} 

\subsection{Proposed Method}\label{sec:network}
\autoref{eq:inverse_problem} can be solved by introducing an auxiliary variable $\vec{x}=q\left(\vec{p}\right)$ and applying variable splitting \cite{Zimmermann2024, Kofler2024_2, Xin2022, Afonso2010}. \autoref{eq:inverse_problem} then becomes
\begin{equation}
      \text{arg}\min_{\vec{x},\vec{p}}||\vec{SFC}\vec{x}-\vec{k}||^{2}_{2}+\lambda||\vec{x}-q\left(\vec{p}\right)||^{2}_{2}+\mathcal{R}\left(\vec{p}\right).
    \label{eq:inverse_problem_splitting}
\end{equation}
For the solution of this minimization, two sub-problems are solved alternatingly for $J$ iterations \cite{Afonso2010}. The first sub-problem is solved to determine the auxiliary variables $\vec{x}_{j+1}$:
\begin{equation}
    \vec{x}_{j+1}= \text{arg} \min_{\vec{x}}||\vec{SFC}\vec{x}-\vec{k}||^{2}_{2}+\lambda||\vec{x}-q(\vec{p}_j)||^{2}_{2}
    \label{eq:first_problem}
\end{equation}
The second sub-problem is solved to update the quantitative parameters $\vec{p}_{j+1}$:

\begin{equation}
    \vec{p}_{j+1}= \text{arg} \min_{\vec{p}}
    \lambda||\vec{x}_{j+1}-q(\vec{p})||^{2}_{2}+\mathcal{R\left(\vec{p}\right)}\,\,\,\,\,\,\,\,\,\,\,\,
    \label{eq:second_problem}
\end{equation}
where $\lambda\in\mathbb{R}$ is a learnable regularization parameter.
\subsubsection{Update of image data}
\label{sec:dc_step}
The first sub-problem, stated in \autoref{eq:first_problem}, can be rewritten to
\begin{equation}
    \min_{\vec{x}}  \frac{1}{2}\left(||\vec{SFC}\vec{x}-\vec{k}||^{2}_{2}+\lambda||\vec{x}-q\left(\vec{p}_j\right)||^{2}_{2}\right).
    \label{eq:first_problem_function}
\end{equation}
Using $\vec{S}^H\vec{S}=\vec{S}$, the condition for minimum is 
\begin{equation}
    \Rightarrow \vec{C}^{H}\vec{F}^{H}\vec{SFCx}_{j+1}+\lambda\vec{x}_{j+1} = \vec{C}^{H}\vec{F}^{H}\vec{S}^H\vec{k}+\lambda q\left(\vec{p}_j\right).
    \label{eq:d_derivative_mod}
\end{equation}
With $\vec{A}=\vec{C}^{H}\vec{F}^{H}\vec{SFC}+\lambda\vec{I}$ and $\vec{b} = \vec{C}^{H}\vec{F}^{H}\vec{S}^H\vec{k}+\lambda q(\vec{p}_j)$ \autoref{eq:d_derivative_mod} becomes $\vec{Ax}=\vec{b}$. $\vec{I}$ stands for the identity matrix. A conjugate gradient algorithm with a tolerance of 1e-4 and unrolled differentiation can then be used to solve \autoref{eq:d_derivative_mod} for $\vec{x}_{j\mathbin{+}1}$ \cite{Nocedal2006}.

\subsubsection{Update of quantitative parameters}
Motivated by the variable-splitting formulation, we approximate the
second subproblem in \autoref{eq:second_problem} using two learned
steps: parameter estimation and spatial regularization. First, the
quantitative parameters $\vec{h}_{j+1}$ are calculated as follows 

\begin{equation}
    \vec{h}_{j+1} = \text{arg} \min_{p}\lambda ||q(\vec{p})-\vec{x}_{j+1}||^{2}_{2} 
    \label{eq:2nd_problem_g}
\end{equation}
This is a non-linear, often non-convex problem with non-unique solutions. We approximate this step by a neural network, denoted here as the \textit{parameter sub-network} $H_{\theta}$:
\begin{equation}
    \vec{h}_{j+1} \approx H_{\theta}(\vec{x}_{j+1})
    \label{eq:parameter_network}
\end{equation}
Subsequently,  for the purpose of regularization, the proximal operator \cite{Chang2023}  of the regularizer $\text{Prox}_{\mathcal{R}:\lambda}$ is employed to solve for the quantitative parameters $\vec{p}_{j+1}$:
\begin{equation}
    \vec{p}_{j+1} =  \text{Prox}_{\mathcal{R}:\lambda}(\vec{h}_{j+1})=\text{arg} \min_{p} \mathcal{R}(\vec{p})+\frac{\lambda}{2}||\vec{p}-\vec{h}_{j+1}||^{2}_{2} 
    \label{eq:2nd_problem_p}
\end{equation}
This step can be replaced by a neural network, denoted here as the \textit{denoiser sub-network} $P_\chi$. Since in \autoref{eq:2nd_problem_p} the parameters $\vec{h}_{j}$ are used, the entire second problem can be stated as an interplay of both neural networks $H_\theta$ and $P_\chi$. Hence we can write for $\vec{p}_{j}$:

\begin{equation}
    \vec{p}_{j+1} \approx P_{\chi}(\vec{h}_{j+1}) \approx P_{\chi}(H_{\theta}(\vec{x}_{j+1}))
    \label{eq:denoiser_network}
\end{equation}
\subsubsection{Signal simulation}
Finally, as a differentiable and numerically faster surrogate to complex MR signal model, we use a pre-trained neural network \cite{Chen2020}. The signal model $q$ can be replaced by a neural network denoted here as the \textit{signal sub-network} $Q_\phi$:\begin{equation}
    q(\vec{p}_{j+1}) \approx Q_{\phi}(\vec{p}_{j+1})
    \label{eq:signal_network}
\end{equation}
This network can be exchanged according to the required signal model and quantitative parameters. By doing so, greater flexibility towards various signal models for different qMRI applications can be achieved.

\subsubsection{Network structure}
\begin{figure*}[!htt]
    \centering
    \adjustbox{trim=0 0 0  0}{
    \includegraphics[width=1\linewidth]{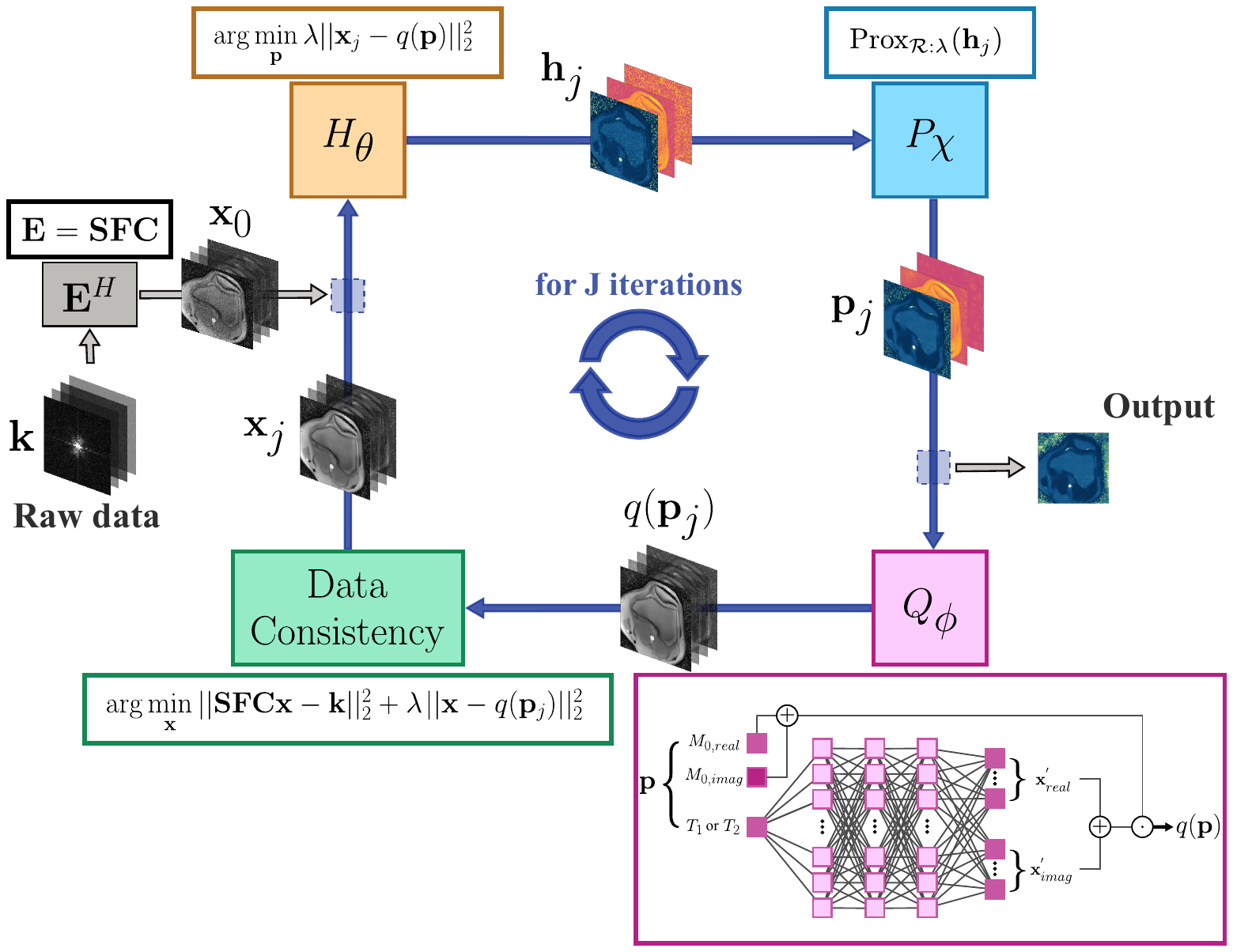}}
    \caption{\textbf{Structure of the proposed approach to solve the qMRI problem.} The approach consists of four components: three sub-networks and a data consistency step. The sub-network $H_\theta$ predicts quantitative parameters from a series of signal images. $P_\chi$ denoises these parameters and the sub-network $Q_\phi$ outputs a series of signal images based on the denoised parameters. Finally, these signal images are forwarded to the data consistency step and used for regularization. The network iterates $J$ times over the four components. The output of the network is the quantitative map resulting from the sub-network $P_\chi$.}
    \label{fig:network_overview}
\end{figure*}

The method explained in this section is implemented into a modular network structure that is shown in \autoref{fig:network_overview}. First, the k-space data $\vec{k}$ is transformed into image data $\vec{x}_0$ via the adjoint of the encoding operator $\vec{E}^{H}$. These signal images $\vec{x}_0$ are then forwarded to the parameter sub-network $H_\theta$ to estimate the corresponding quantitative MR maps $\vec{h}_j$. The denoiser sub-network $P_\chi$ then denoises these maps and outputs $\vec{p}_j$. The two latter steps describe the update of the quantitative parameters described in \autoref{eq:denoiser_network}. The parameters are subsequently passed on to the signal sub-network $Q_\phi$ which outputs the corresponding signal images $q(\vec{p})$ that serve as a regularizer for the following data consistency step. In this step the image data $\vec{x}_j$ is updated.  This loop is repeated for $J$ iterations. The $J$-th output of the denoiser sub-network $\vec{p}_J$ is the final quantitative map estimated by the proposed network. 
\subsection{Implementation}\label{sec:experiments}
The proposed network consists of three different sub-networks $H_\theta$, $P_\chi$ and $Q_\phi$ . In the following the structure and training of each sub-network is presented.  \\
\subsubsection{Sub-network $H_{\theta}$}
Starting with $H_\theta$, this sub-network is a multilayer perceptron (MLP) \cite{Chan2023}. It consists of four ReLU-activated hidden layers of size 256. The input of $H_\theta$ are signal images for a number of time points. The real and imaginary parts of the signals are concatenated and scaled by the maximum value of the absolute signal values. Hence, the input size of this sub-network is equal to $2\cdot N_a$.  Since all batch samples are merged into one dimension, $H_\theta$ works on single pixels and is independent of any anatomy or image structure. Regarding the output of this sub-network, the number of quantitative parameters $\vec{h}_j$ determines the size of the output layer. A softplus function is used for the (bio-)physical parameters expected to be positive (e.g. relaxation times).\\

\subsubsection{Sub-network $P_\chi$}
The sub-network $P_\chi$ consists of a U-Net \cite{Ronneberger2015} made up of three encoding and three decoding layers. The maximum number of channels reached during the application of this network is 512 with $15\times15$ pixels as the lowest resolution. The U-Net is applied separately to the different parameter maps and the input is scaled by the standard deviation of the absolute values of the quantitative parameter maps. This network consists of around 5.7 million trainable parameters. In addition to the U-Net the concept of residual networks \cite{He2016} is employed, so that the quantitative output $h_{denoised}$ of $P_\chi$ is added to the noisy input $\vec{h}_j$ of $P_\chi$: 
\begin{equation}
    \vec{p}_{j}  = \vec{h}_{j} + \gamma \cdot \vec{h}_{j, denoised}
    \label{eq:denoiser_residual}
\end{equation}
$\gamma\in\mathbb{R}$ is a scaling parameter that is optimized during the training of the proposed network. A softplus function is applied to $\gamma$ before every usage. \\

\subsubsection{Sub-network $Q_\phi$ }\label{sec:Q}
The last sub-network $Q_\phi$ is again a MLP with ReLU-activation functions. An overview of this sub-network is given in \autoref{fig:network_overview}. It consists of three hidden layers of size 128. The size of the output layer is determined by the number of time points in the respective signal model. The input of the MLP are flattened quantitative parameters. $Q_\phi$ only works on single pixels, similar to $H_\theta$. The output of the last layer consists of the real and imaginary components of the signal. Finally, these components are scaled with the quantitative parameter $M_0$ to obtain the complex signal $q(\vec{p})$ for each pixel over all time points. $Q_\phi$ is the only component of the proposed network that is trained prior to the actual training of the network.\\ 
 
\subsubsection{Training}

The proposed network, shown in \autoref{fig:network_overview}, is trained on Dataset 1. Only the sub-network $Q_\phi$ is pre-trained and then kept fixed during the training of the proposed approach, as mentioned in section \ref{sec:Q}. This pre-training is based on the mono-exponential decay signal model presented in section \ref{sec:training_data}. We focus here on a simple signal model but it could easily be replaced with a more complex signal model, such as extended phase graphs. The signal sub-network $Q_\phi$ outputs signal images based on the incoming parameter maps and the selected signal model. For the loss calculation the mean squared error (MSE) between the signal images corresponding to the input quantitative parameters and the output signal images is calculated, where the imaginary and real signal parts are handled separately. \\
During the training of the proposed network the parameter sub-network $H_\theta$ is trained as part of the larger network. $H_\theta$ is only separately trained for the application of the proposed network on Dataset 2 and 3. During the training of $H_\theta$ the signal input is created using a signal model on random quantitative parameters. The loss for the training of this sub-network is equal to the MSE of the quantitative parameters corresponding to the input signal and the quantitative parameters being output by $H_\theta$, where the imaginary and real parts of $M_0$ are handled separately.\\
The sub-network $P_\chi$ is not trained separately but is trained during the training of the proposed approach. For the application on different datasets after training, this sub-network does not have to be re-trained since it works independently of the given signal model or the required quantitative parameters.\\
During the training of the proposed network, the loss $\mathcal{L}$ is calculated by calculating the MSE of different parameters.
$\mathcal{L}$ considers the final and intermediate outputs of the parameter sub-network $H_\theta$, namely $\vec{h}_{j}$, and the  final and intermediate outputs of the denoiser sub-network $P_\chi$, therefore $\vec{p}_{j}$. These parameters are compared to the corresponding ground truth parameter maps $\vec{p}_{GT}$.  For the loss calculation the concepts of supervised training \cite{Kofler2026} and deep supervision \cite{Lee2015} were employed. The per-sample loss $\mathcal{L}$ was calculated as
\begin{equation}
    \mathcal{L} = \frac{1}{J}\sum_{j=1
}^J||\mathbf{M}\odot(\vec{h}_{j}-\vec{p}_{GT})||^2_2+||\mathbf{M}\odot(\vec{p}_{j}-\vec{p}_{GT})||^2_2 \,,
\end{equation}
with $\mathbf{M}$ a sample specific mask keeping only anatomical structures. 
\section{Experiments}
 
\subsection{Training and evaluation data}\label{sec:training_data}
Dataset 1 was the dataset used for the training and evaluation of the network. It consisted of simulated k-space data obtained from $M_{0}$- and $T_2$-maps of knee joints from ten different subjects acquired on a Siemens MAGNETOM Free.Max scanner, with a magnetic field strength of $0.55\,$T. These quantitative maps were used as an input to the signal model $q$ in \autoref{eq:inverse_problem} to create training data pairs ($k$, $p_{true}$). Truncated gaussian noise with a mean value of $\mu=0$ and a standard deviation of $\sigma=0.05$ was added to $k$. A mono-exponential decay model was used as $q$ with echo times spaced equally between 16\,ms and 160\,ms. For this work 2D scans covering the knee with a matrix size of $80\times80$ pixels were used.   \\
The provided $M_0$-maps were real valued and a random phase component was added to the $M_0$-maps to be used in the training. The dataset was augmented with random rotation and flipping of the quantitative maps. In total 2400 slices were given in this dataset for 80 slices in three different directions for ten different subjects. For this approach the split of training-/validation-/test-data was 70/10/20.
\subsection{Application data}\label{sec:application_data}
 Dataset 2 was acquired from an MRI Halbach scanner with a magnetic field strength of $72\,$mT and consisted of single-coil k-space data of the knee of five different subjects. \footnote{\url{https://zenodo.org/records/18460613}} The scans were obtained with a 3D STIR sequence for $T_1$-mapping with the following inversion times: 0, 40, 80, 120, 160, 200, 500\,ms. The field-of-view  and the matrix size used in this case were $240\times180\times160\,$mm$^{3}$ and $160\times120\times32$. \\
Dataset 3 consisted of multi-coil k-space data of the brain of ten different subjects acquired with a $T_2$ GraSE (EPI) Sequence  on a Philips Ingenia Ambitions X scanner that was ramped down to a magnetic field strength of $0.6\,$T. \footnote{\url{https://zenodo.org/records/18847561}} The coil sensitivity maps were calculated with the Inati method \cite{Inati2013,Inati2014}. The field-of-view for the scans was $230\times180\,$mm$^{2}$ and the matrix size employed was $119\times152$ and three slices. The echo times for the acquisitions were 21, 32, 43, 54, 64, 75, 86, 96, 107 and 118 \,ms. Both datasets are publicly available \cite{zenodo_spain,zenodo_switzerland}.

\subsection{Training settings}
For the training of the proposed network an AdamW-optimizer \cite{Loshchilov2019} was used. Furthermore gradient clipping was employed. The learning rate was 0.0001 and the batch size was 8. In addition to the parameters of the parameter sub-network $H_\theta$ and the denoiser sub-network $P_\chi$, the residual scale factor $\gamma$ and the regularizing factor $\lambda$ were optimized during the training of the proposed network.  $\gamma$ was set to 0.01 and $\lambda$ was set to 0.5 at the beginning of the training.  Before every usage of these two parameters, a softplus function was applied to them. The number of iterations $J$ for this training was set to 4. In total 5,859,078 parameters were trained.\\
 For the pre-training of $Q_\phi$ the AdamW-optimizer was employed. The learning rate was 0.0001 and the batch size was 64. 100,000 uniformly distributed random $M_0$-, $T_1$- and $T_2$-values were used. Either $T_1$- or $T_2$-values were used, depending on the relevant quantitative parameters map that was reconstructed by the proposed network. The random $M_{0,real}$- and $M_{0,imag}$-values both ranged from -1 to 1. The $T_1$-values and $T_2$-values ranged both from 5\,ms to 500\,ms. \\
 The sub-network $H_\theta$ was only separately trained when the proposed approach was applied to Dataset 2 and 3. The same optimizer, learning rate and batch size were the same as for the pre-training of $Q_\phi$. 100,000 uniformly distributed random $M_0$, $T_1$ or $T_2$-values were used. They ranged between -1 and 1 for the real and imaginary parts of $M_0$ and between 5\,ms and 500\,ms for $T_1$ and $T_2$. For all datasets  the real and imaginary part of the quantitative parameter $M_0$, namely the proton density, made up two of the outputs of $H_\theta$.

\subsection{Comparison to other methods}\label{sec:comparison}
The proposed network and its application on Dataset 1 was compared to three other reconstruction methods, two classical methods, such as direct reconstruction  and subspace reconstruction based on PCA compression \cite{Pfister2019}, both with subsequent dictionary matching, and to the deep learning based reconstruction method \textit{Mantis}  \cite{Liu2019}.\\
For the direct reconstruction method a qualitative image was first reconstructed with an adjoint encoding operator. A dictionary with different signal curves was created. Using the signal model of the mono-exponential decay described in section \ref{sec:training_data}, these signal curves were calculated based on 1000 random $T_2$-values between 5\,ms and 500\,ms. The k-space was transformed into signal images and subsequently matched to the signal curves in the dictionary to generate the $T_2$-map.\\
For the subspace method first a dictionary of different signal curves was created based on the mono-exponential signal model mentioned in section \ref{sec:training_data}. This dictionary was created with 1000 $T_2$-values ranging from 5\,ms to 500\,ms. A PCA-compression operator was then created based on a two-component PCA-compression of these dictionary signal curves. In addition to the PCA-compression the reconstruction problem was solved with the conjugate gradient method and Tikhonov regularization with a regularization parameter of 0.05. Afterwards, the corresponding $T_2$-values were estimated with the method of dictionary matching with the settings used for the PCA compression operator above.\\
For Mantis first qualitative images were reconstructed from k-space data via an adjoint Fourier operator and then a U-Net was applied to the reconstructed absolute signal images to output denoised quantitative parameter maps. For the training of Mantis Dataset 1 was used and augmented by random rotation and flipping. No random phase was added to the $M_0$-values, since the U-net was only applied to the absolute signal. k-space data was obtained as described in section \ref{sec:training_data}. Losses of quantitative parameters and losses in k-space were considered, where the scale factor for the latter was set to $\lambda=0.1$.\\
These methods were compared visually, as well as quantitatively by calculating the mean absolute difference of the masked ground truth $T_2$-maps to the masked $T_2$-maps of the different methods. This was done for all 480 slices of the test dataset of Dataset 1.

\subsection{Generalization Study}
The generalizability of the proposed approach was evaluated on Dataset 2 and Dataset 3. To do so, the sub-networks $H_\theta$ and $Q_\phi$ had to be re-trained on purely synthetic signal curves independent of actual data. Compared to the previous application on Dataset 1, only the signal model and the parameters had to be adjusted. Akin to Dataset 1, for Dataset 3 the signal model was also the mono-exponential decay signal model but with different decay times, as mentioned in section \ref{sec:application_data}. Hence, for the application on Dataset 3  the relevant quantitative parameters were also $M_0$ and $T_2$.  For Dataset 2, the parameters were $M_0$ and $T_1$ and the signal model was changed to an inversion recovery signal model with the inversion times mentioned in section \ref{sec:application_data}.

\section{Results}\label{sec:results}
\subsection{Application on evaluation dataset}

\begin{figure*}[!ht]
    \centering
    \adjustbox{trim=1cm 2cm 0  2cm,clip}{
    \includegraphics[width=1\linewidth]{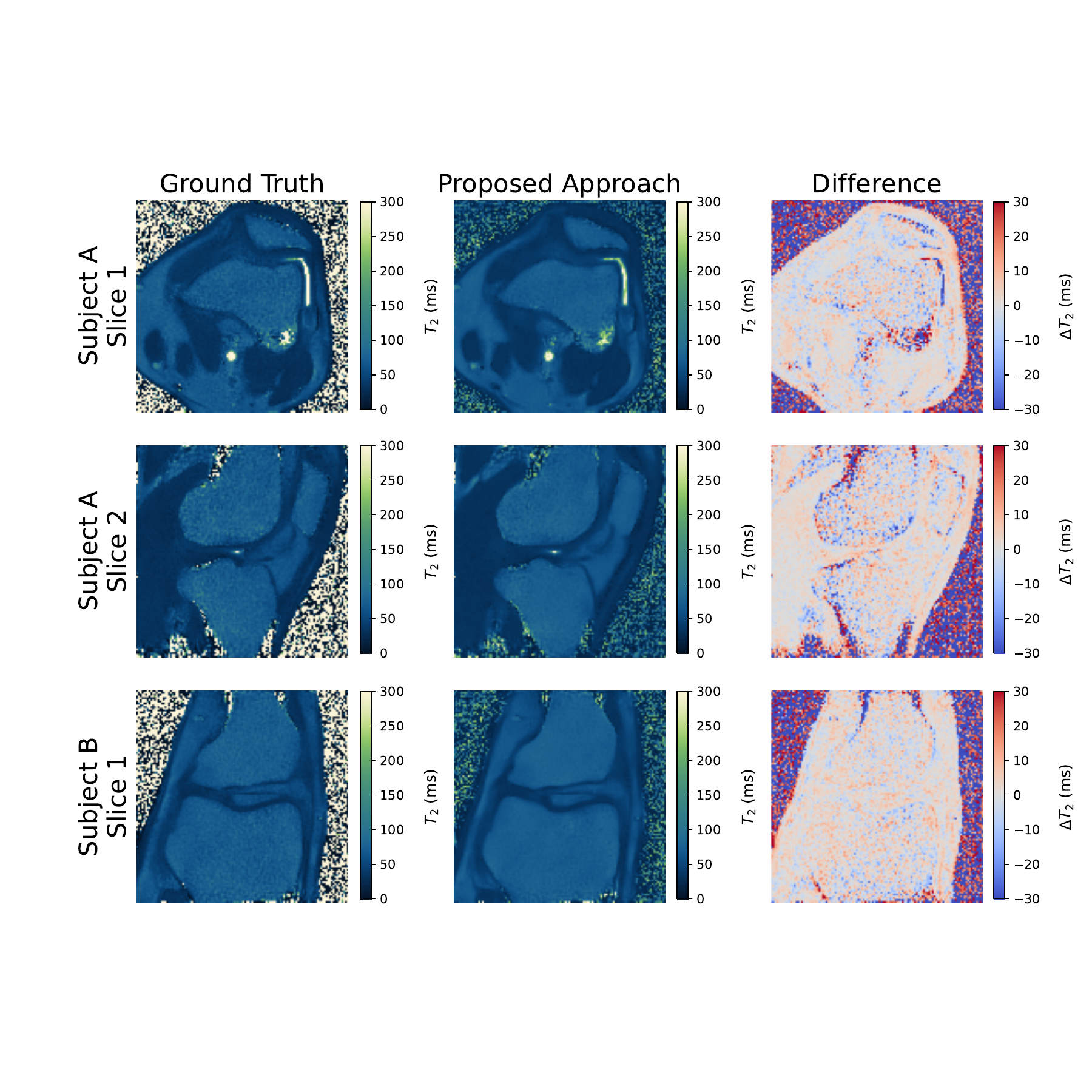}}
    \caption{\textbf{Application of proposed method on Dataset 1.} Resulting $T_2$-maps estimated with the proposed approach for three different slices from Dataset 1. The differences to the corresponding ground truth maps are shown in the far right column.} 
    \label{fig:application}
\end{figure*}
The ground truth $T_2$-maps and the corresponding maps resulting from the proposed approach can be seen for three exemplary slices in \autoref{fig:application}. As can be seen in these maps, the predicted $T_2$-values by the proposed method match the ground truth values very well. For the presented slices many pixels show differences up to $10\,$ms to ground truth.

\subsection{Comparison to other methods}
\begin{figure*}[!htt]
    \centering
    \adjustbox{trim=1cm 1.5cm 0  1cm,clip}{
    \includegraphics[width=1.1\linewidth]{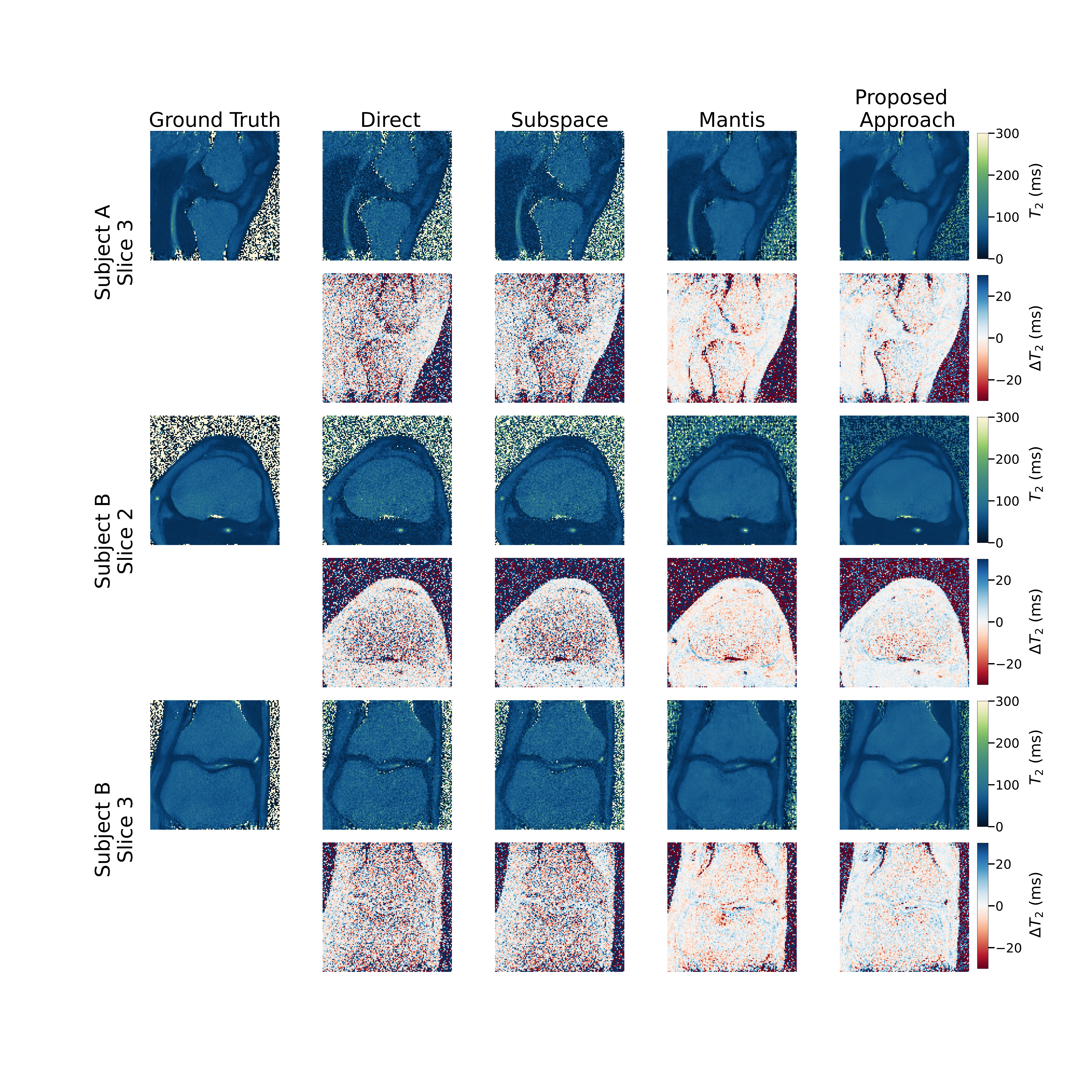}}
    \caption{\textbf{Comparison of proposed approach to other methods.} $T_2$-maps estimated with direct reconstruction, the subspace method, the Mantis network and the proposed approach for three different slices from Dataset 1. The corresponding difference maps to the ground truth $T_2$-maps are shown underneath.}
    \label{fig:comparison}
\end{figure*}

\begin{figure}
 \centering
    \adjustbox{trim=0.2cm 0cm 0  0cm,clip}{
    \includegraphics[width=0.7\linewidth]{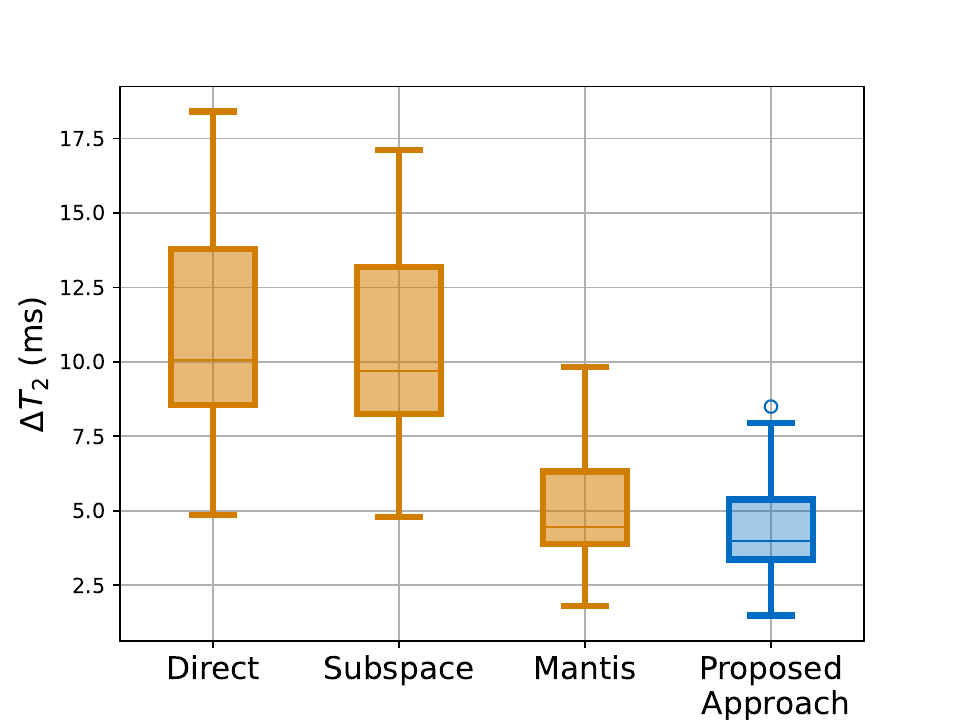}}
    
    \centering
         \caption{\textbf{Boxplot of mean absolute differences for all four methods.} Mean absolute differences between the masked ground truth $T_2$-maps and the  masked $T_2$-maps estimated with direct reconstruction, the subspace method, the Mantis network and the proposed method. The boxplots show the respective results for all 480 slices in the test dataset of Dataset 1.   }
         \label{fig:comparison_boxplot}

\end{figure}

The comparison of different reconstruction methods is shown in \autoref{fig:comparison}. The proposed method shows better $T_2$-maps than the classical approaches and also the deep learning reference method (Mantis). This is also confirmed by \autoref{fig:comparison_boxplot} which shows the quantitative evaluation over the entire test dataset. For the proposed approach the median absolute difference for the test dataset is lower than $4\,$ms while the maximum absolute difference is $8.50\,$ms  for one slice.  The mean absolute difference of the proposed method is $4.31\pm1.30\,$ms compared to $11.01\pm3.07\,$ms for the direct method, $10.54\pm2.84$\,ms for the subspace method and $4.98\pm1.53\,$ms for Mantis. \\

\begin{figure*}[!htt]
    
\end{figure*}

\subsection{Generalization study}
\begin{figure*}[!ht]
    \centering
    \adjustbox{trim=1cm 2cm 0  1cm,clip}{
    \includegraphics[width=1\linewidth]{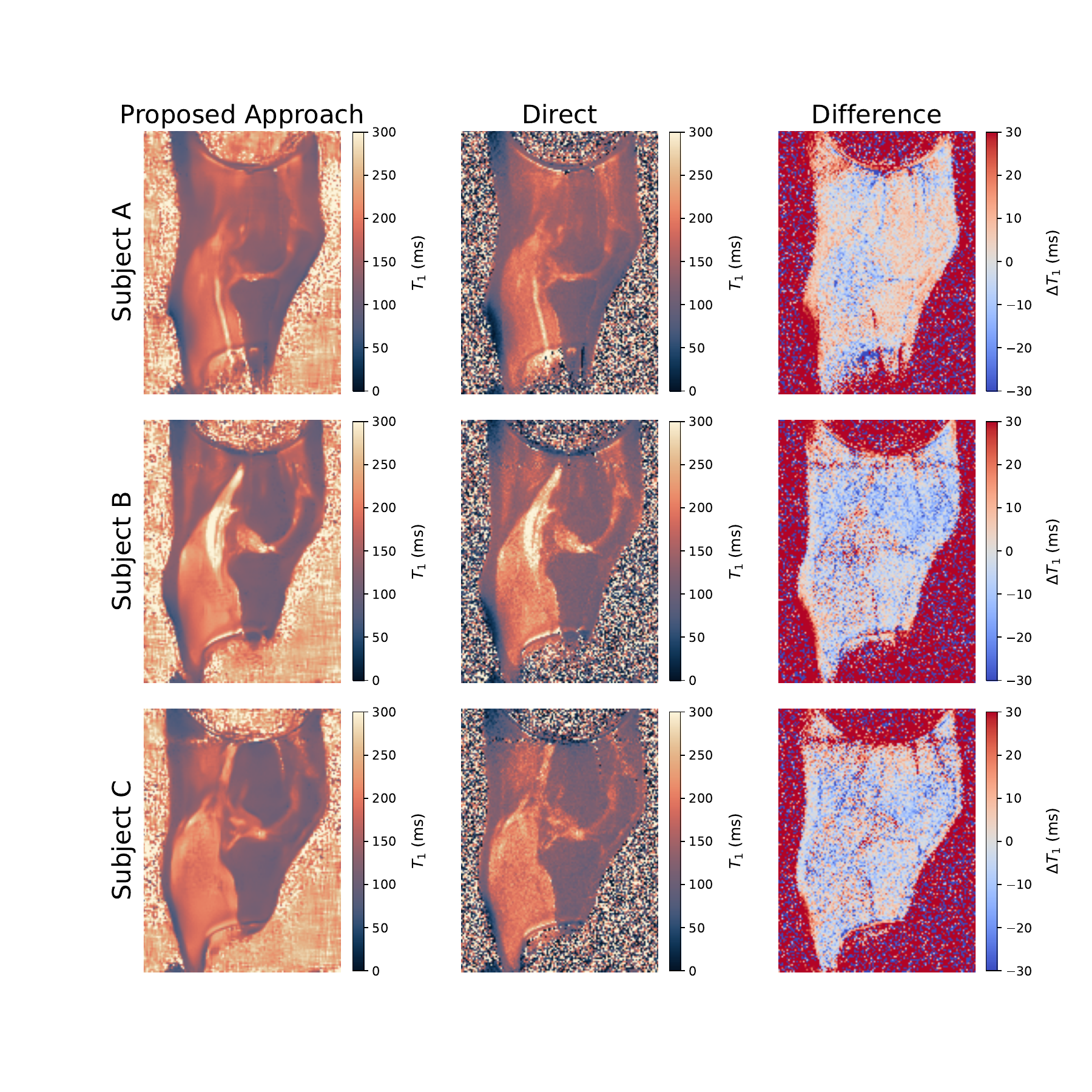}}
    \caption{\textbf{Application of proposed method on Dataset 2.} Resulting $T_1$-maps estimated with the proposed approach for three different slices from Dataset 2 (72\,mT). The corresponding $T_1$-maps reconstructed with the direct reconstruction approach are shown, as well as the difference maps to the outputs of the proposed method.   } 
    \label{fig:application_spain}
\end{figure*}

\begin{figure*}[!ht]
    \centering
    \adjustbox{trim=1cm 2cm 0  1cm,clip}{
    \includegraphics[width=1\linewidth]{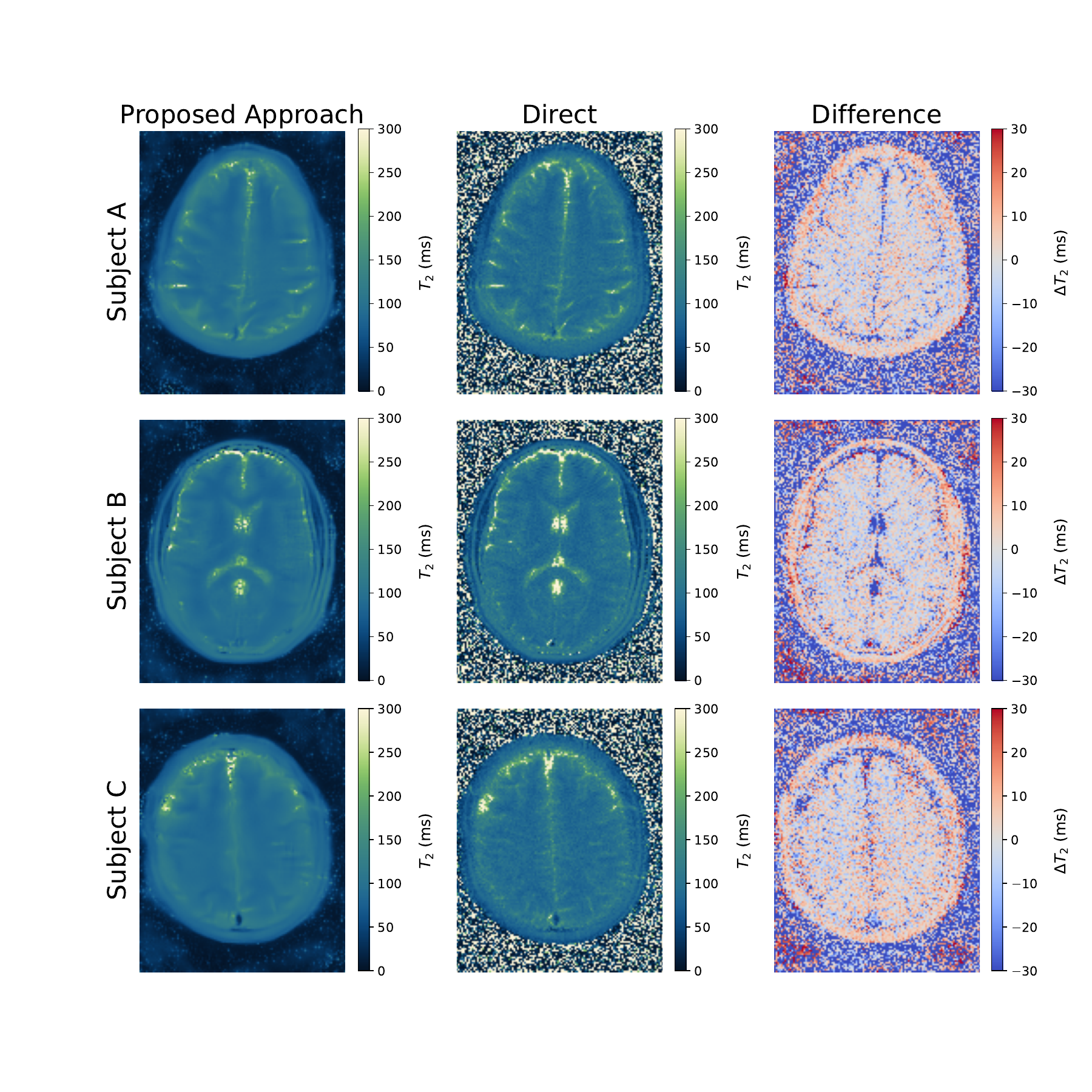}}
    
    \caption{\textbf{Application of proposed method on Dataset 3.} Resulting $T_2$-maps estimated with the proposed approach for three different slices from Dataset 3 (0.6\,T). The corresponding $T_2$-maps reconstructed with the direct reconstruction approach are shown, as well as the difference maps to the outputs of the proposed method.  } 
    \label{fig:application_switz}
\end{figure*}

 The application of the proposed approach on three different slices of Dataset 2 are shown in \autoref{fig:application_spain}. The resulting $T_2$-maps for the application of the proposed method on three different slices of Dataset 3 are shown in \autoref{fig:application_switz}.
\clearpage
\section{Discussion}\label{sec:discussion}
The proposed approach yields accurate quantitative MR maps, directly from acquired k-space data, enabling fast applications in low-field MRI. Applying the presented method on Dataset 1 yielded $T_2$-maps with mean absolute differences under 8.6\,ms.  The method outperformed other methods such as the Mantis network, the subspace or the direct reconstruction method. Nevertheless, the corresponding difference map in \autoref{fig:comparison} still shows some differences in the outer bone area. This could potentially be improved by increasing the number of encoding and decoding layers in the U-Net of the denoiser sub-network $P_\chi$.  \\

We could demonstrate that the proposed network is highly flexible in terms of required signal model, anatomy and magnetic field strength while still enforcing data consistency. The network could be trained for $T_2$-mapping of the knee at 0.55\,T and was then applied to $T_2$-mapping of the brain at 0.6\,T and $T_1$-mapping of the knee at 72\,mT.
This independence of the approach to the anatomy is mostly due to the sub-networks $H_\theta$ and $Q_\phi$, responsible for the parameter and signal estimation, which are applied pixel-wise. The approach could be adapted to different signal models by simply re-training the two sub-networks $H_\theta$ and $Q_\phi$ on completely synthetic data. Therefore no additional acquisition of training data was necessary. The proposed method can be adapted to other signal models and thus be used for different qMRI techniques. The signal models are incorporated as trained networks. Therefore, also complex signal models such as extended phases graphs, can be used without leading to a significant increase in training time and which would, for instance, not be possible with a deep learning network such as PINQI \cite{Zimmermann2024}. We also showed that it is independent of the magnetic field strength that was used to acquire the data. Although it was trained on Dataset 1, which was acquired at a field strength of 0.55\,T, we were able to show that it can still be applied to data acquired at 0.6\,T or even 72\,mT. This is especially advantageous since there is a lack of training data especially for field strengths below 0.1\,T. Additionally we were able to apply the proposed approach to multi-coil k-space data, namely Dataset 3, even though the network was only trained with single-coil data. \\

A limitation of this work was the missing ground truth for Dataset 2 and 3 which impeded a quantitative evaluation, as differences from the direct reconstruction quantify agreement rather than accuracy. Nevertheless, the difference images in \autoref{fig:application_spain} and \autoref{fig:application_switz} do not show any strong anatomical features, but are mainly noise-like, hinting at a successful transfer without induced hallucinations.
\section{Conclusion}\label{sec:conclusion}
In this work a deep learning-based approach to obtain accurate quantitative maps directly from acquired k-space data for low-field MRI was presented. It can be applied to datasets that differ in terms of quantitative MR parameters, anatomy and magnetic field strength. We demonstrated that it handles noisy data successfully while maintaining high accuracy. The presented network is an efficient and flexible solution to reconstruct quantitative low-field MR maps directly from k-space data.

\clearpage 

\bibliography{references} 
\bibliographystyle{sciencemag}

\paragraph*{Funding:}
This research was supported by the project  22HLT02 A4IM which has received funding from the European Partnership on Metrology, co-financed from the European Union’s  Horizon Europe Research and Innovation Programme and by the  Participating States.

\paragraph*{Data, code and materials availability:}
The presented method and the other methods used for comparison were implemented with the PyTorch-based MRI reconstruction package \textit{MRpro} \cite{Zimmermann2026}. Dataset 2 and 3 are publicly available \cite{zenodo_spain,zenodo_switzerland}.

\end{document}